\documentclass[twocolumn,aps,prl,showkeys,superscriptaddress,hyperref,letterpaper]{revtex4-1}
\usepackage[latin9]{inputenc}
\usepackage{verbatim}
\usepackage{amsmath}
\usepackage{amssymb}
\usepackage{graphicx}
\usepackage{color}

\usepackage[unicode=true,
 bookmarks=true,bookmarksnumbered=false,bookmarksopen=false,
 breaklinks=false,pdfborder={0 0 0},pdfborderstyle={},backref=false,colorlinks=true,]
 {hyperref}
\hypersetup{linkcolor=blue,citecolor=blue,urlcolor=blue}

\makeatletter

\newcommand{\hc}{\mathrm{H.c.}}

\usepackage{bm}\renewcommand{\mathbf}{\bm}

\makeatother

\begin{document}

\title{Quantum phase diagram of Shiba impurities from bosonization}


\author{Tom\'as Bortolin}

\affiliation{Instituto de F\'{\i}sica La Plata - CONICET and Departamento de F\'isica, Universidad Nacional de La Plata, cc 67, 1900 La Plata, Argentina.}

\author{An\'{\i}bal Iucci}

\affiliation{Instituto de F\'{\i}sica La Plata - CONICET and Departamento de F\'isica, Universidad Nacional de La Plata, cc 67, 1900 La Plata, Argentina.}

\author{Alejandro M. Lobos}\email{alejandro.martin.lobos@gmail.com}

\affiliation{Facultad de Ciencias Exactas y Naturales, Universidad Nacional de Cuyo and CONICET, 5500 Mendoza, Argentina}

\begin{abstract}
A characteristic feature of Shiba impurities is the existence of a spin- and parity-changing quantum phase transition (known as ``0-$\pi$'' transition) which has been observed in scanning tunneling microscopy (STM) and transport experiments. Using the Abelian bosonization technique, here we analyze the ground-state properties and the quantum phase diagram of a classical (i.e., Ising-like) Shiba impurity. In particular, we analyze the cases of an impurity in a three- and a one-dimensional superconductor. Within the bosonization framework, the ground-state properties are determined by simple  soliton-like solutions of the classical equations of motion of the bosonic fields, whose topological charge is related to the spin and parity quantum numbers. Our results indicate that the quantum phase diagram of the superconductor can be strongly affected by geometrical and dimensional effects. Exploiting this fact, in the one-dimensional case we propose an experimental superconducting nanodevice in which  a novel parity-preserving, spin-changing ``0-0'' transition is predicted.
\end{abstract}

\pacs{PACS ACA}

\maketitle

\emph{Introduction.}---
Yu-Shiba-Rusinov (or simply Shiba) states are localized subgap states arising in a superconductor due to the local pair-breaking processes induced by a magnetic impurity~\cite{Yu1965,Shiba1968,Rusinov1968, Balatsky2006}, or by a quantum dot in superconductor-hybrid nanodevices, where they are more commonly known as ``Andreev bound states''~\cite{DeFranceschi2010, Deacon2010, MartinRodero2011, Maurand2012}. Recently, these systems have become the focus of intensive research due to the prediction that Majorana zero-modes (i.e., non-Abelian
quasiparticles with potential uses in fault-tolerant topological quantum
computation~\cite{Nayak2008}) could be realized in a linear chain of magnetic impurities deposited on top of a superconductor. In these systems,  the Shiba states can overlap and form a  topological ``Shiba band''~\cite{Nadj-Perge2013,Klinovaja2013,Braunecker2013,Pientka2013} that mimics the physics of the Kitaev one-dimensional model~\cite{Kitaev2001}.
Subsequent STM studies of chains of Fe atoms deposited on top of clean Pb surfaces have shown compelling evidence for the Majorana
scenario~\cite{Nadj-Perge2014,Pawlak2016,Ruby2015}, generating a lot of excitement.  

A salient feature of Shiba systems is the existence of an experimentally accessible quantum phase transition,  known as the ``0-$\pi$ transition'', determined by the position of the level inside the superconducting gap: if the Shiba state falls below the Fermi energy, it becomes occupied and the nature of the collective ground state changes from an even-parity BCS-like singlet to an odd-parity spin-1/2 doublet~\cite{Sakurai1970}. This  parity- and spin-changing transition has been experimentally observed  both in adatom/superconductor systems with STM techniques~\cite{Franke2011,Bauer2013}, and in superconductor-quantum dot devices via quantum transport experiments~\cite{Deacon2010, Lee2012, Maurand2012, Schindele2014, Island2017}. 

Recent progress in nanofabrication has allowed the realization of ultrathin superconducting epitaxial nanostructures [i.e., one-dimensional nanowires (1DNWs)] using the proximity effect~\cite{Krogstrup2015,Taupin2016}. These advances pave the way for novel technological applications, and can potentially enable the study of Shiba states in systems with reduced dimensionality. A question which naturally arises in this context is whether the different dimensionality or geometrical properties of the superconductor can qualitatively modify the properties of Shiba systems and the above-mentioned phase diagram. For instance, scattering processes which are crucial in the one-dimensional (1D) geometry can be profoundly affected by interactions \cite{Fisher1992a,Kane1992}. In a 1D system (such as a proximitized 1DNW), the enhanced effect of correlations might indeed change the properties of a superconductor, thus modifying the features of the induced quantum phases.

In this work we implement the Abelian bosonization formalism to study the zero-temperature phase diagram of a Shiba impurity, both in a 1D and in a three dimensional (3D) superconductor. Interestingly, within the bosonic representation the ground-state properties in both geometries can be studied in an unified way. In particular for the 1D case, we propose a novel experimental device based on proximitized  semiconducting 1DNWs with a ferromagnetic (FM) nanocontact grown ontop to induce a controllable Shiba state. We predict that this device could host a novel parity-conserving,spin-changing ``0-0'' transition. Our results might have impact in recent theoretical and experimental developments where Shiba states have been observed, and in recent works where superconductivity is induced in semiconducting nanowires by the means of proximity effect.

\emph{Theoretical model.}---
We begin by analyzing the 1D geometry. We describe a proximitized single-channel superconducting 1DNW of length $2L_\text{W}$ with the Hamiltonian $H=H_{0}+H_{\Delta}+H^\text{1D}_{M}$, where
\begin{equation}
H_0=\int_{-L_\text{W}}^{L_\text{W}}  dx \left[\sum_\sigma \Psi_\sigma^\dagger\left( -\frac{\partial_x^2}{2m}\right)\Psi_\sigma + U \Psi_\uparrow^\dagger \Psi_\uparrow \Psi_\downarrow^\dagger \Psi_\downarrow\right]
\end{equation}
 describes the (interacting) semiconductor wire. Here, $\Psi_\sigma$ anhilates a fermion with spin $\sigma=\uparrow,\downarrow$,  and $U>0$ is a repulsive Hubbard-type interaction parameter (here we have used $\hbar=1$). Linearization of the 1DNW spectrum in a region of width $2\Lambda$ around the Fermi energy allows to write the Fermi fields as $\Psi_{\sigma}\left(x\right) =e^{ik_{F}x}\psi_{L\sigma}\left(x\right)+e^{-ik_{F}x}\psi_{R\sigma}\left(x\right)$, where $k_F$ is the Fermi wavevector, and $\psi_{r\sigma}$ ($r=L,R$) are left/right moving chiral fields slowly-varying  on the scale $k_F^{-1}$.
We now introduce the bosonization formalism~\cite{Giamarchi2004, Gogolin1988}, and represent the chiral fermion fields as $\psi_{r\sigma}=F_{r\sigma}e^{-ir\phi_{r\sigma}}/\sqrt{2\pi a}$ where $\phi_{r\sigma}$ are chiral bosonic fields obeying the Kac-Moody commutation relations $[\phi_{r\sigma}(x),\phi_{r'\sigma'}(x')]=i\pi r\delta_{r,r',}\delta_{\sigma,\sigma'}\text{sgn}\left(x-x'\right)$, $a\sim k_F^{-1}$ is the short-distance cutoff of the continuum theory, and $F_{r\sigma}$ are standard Klein factors which ensure the proper anticommutation relations of the Fermi fields. 

It is customary to introduce the new fields $\phi_{r\sigma}=\frac{1}{2}[\phi_{c}+r\theta_{c}+\sigma(\phi_{s}+r\theta_{s})]$, where $c$($s$) refers to  charge- (spin-) type operators, and where the convention $r=+(-)$ on the r.h.s has been used for the $R(L)$ branch, and similarly, $\sigma=+(-)$ for $\uparrow (\downarrow)$.
The new fields satisfy canonical commutation relations $[\phi_{\mu}(x),\partial_{y}\theta_{\nu}(y)] =-2i\pi\delta_{\mu\nu}\delta(x-y)$, and  physically, the field $\phi_s$ is related to the spin density operator $\rho_s(x)=\frac{1}{2}\left[\Psi^\dagger_\uparrow\left(x\right)\Psi_\uparrow\left(x\right)-\Psi^\dagger_\downarrow\left(x\right)\Psi_\downarrow\left(x\right)\right]=-\frac{1}{2\pi}\partial_x \phi_s +\frac{1}{4\pi a}\left[e^{i2k_Fx}e^{i\phi_c}\left(e^{i\phi_s}F^\dagger_{R\uparrow}F_{L \uparrow}-e^{-i\phi_s}F^\dagger_{R\downarrow}F_{L \downarrow}\right)+\hc\right]$, while $\theta_c$ is related to the Josephson phase field of the superconductor and to the current density $j\left(x\right)=-2ev_c K_c\partial_x \theta_c\left(x\right)/\pi$. In terms of these fields, the Hamiltonian $H_0$ takes a Luttinger liquid form with decoupled charge and spin bosonic sectors, i.e., $H_0=H_c+H_s$, where
\begin{equation}
H_{\nu}=\frac{v_{\nu}}{4\pi}\int_{-L_\text{W}}^{L_\text{W}}  dx\,\left[\frac{1}{K_{\nu}}(\partial_{x}\phi_{\nu})^{2}+K_{\nu}(\partial_{x}\theta_{\nu})^{2}\right]
\end{equation}
for $\nu=c,s$. The parameter $K_\nu$  encodes the interactions in each sector, and physically controls the decay of the correlation function $\langle \Psi_{r\sigma}\left(x\right)\Psi^\dagger_{r\sigma}\left(x^\prime\right)\rangle\sim \left|x-x^\prime\right|^{-\left(K_c+K_s\right)}$. In our particular case,  $K_{c}=1/\sqrt{1+Ua/\left(\pi v_F\right)}$, and due to the SU(2) symmetry of the model, the value of $K_s$ is constrained to $K_s=1$. On the other hand, $v_c=v_F\sqrt{1+Ua/\left(\pi v_F\right)}$ is the velocity of the 1D charge plasmons, and $v_s=v_F$ is the velocity of the 1D spinons.

Next, the proximity-induced superconducting pairing in the 1DNW  is described by
$H^\text{1D}_\Delta=\Delta \int_{-L_\text{W}}^{L_\text{W}}  dx\,\Psi^\dagger_\uparrow\Psi^\dagger_\downarrow+\hc=\Delta \int_{-L_\text{W}}^{L_\text{W}}  dx\, \left[\psi^\dagger_{L\uparrow}\psi^\dagger_{R\downarrow}+\psi^\dagger_{R\uparrow}\psi^\dagger_{L\downarrow}+\hc\right]$, where we have neglected the rapidly oscillating terms proportional to $e^{\pm i 2k_Fx}$  since they average to zero. Physically, the induced pairing potential $\Delta$ emerges from the integration of the bulk superconductor, and strongly depends on the transparency and disorder of the superconductor/nanowire contact \cite{Takei13_Soft_gap}. In terms of the bosonic fields, the pairing term writes	
$H_{\Delta}  =\frac{2\Delta}{\pi a}\int_{-L_\text{W}}^{L_\text{W}}  dx\, \cos\theta_{c}\left(x\right)\cos\phi_{s}\left(x\right)$. A simple scaling analysis where we change the cutoff $a\rightarrow a\left(1+d\ell\right)$ indicates that $H_{\Delta}$ is a relevant perturbation in the RG sense, i.e., $d\Delta\left(\ell \right)/d\ell=\left(2-1/K_c\right) \Delta\left(\ell\right)$, and flows to strong coupling as $\ell \rightarrow \infty$ when $K_c>1/2$.

Finally, Shiba states emerge due to the presence of a localized magnetic field or impurity in the superconductor. In an ultrathin mesoscopic 1DNW, atomic-sized impurities are hard to control and manipulate experimentally. Therefore, in order to induce a controllable Shiba state, we assume the presence of a FM insulating contact of width $d$ grown ontop of the 1DNW. This Hamiltonian writes
$H^\text{1D}_{M}= V \int_{-L_\text{W}}^{L_\text{W}} dx\ m_z\left(x\right)\rho_s\left(x\right)$, where $V$ is the exchange field induced by the FM contact, and $m_z\left(x\right)$ is its dimensionless magnetization profile, assumed to be oriented along the $z$-axis \cite{Sau2010}. 
For concreteness, we assume a Gaussian profile $m_z\left(x\right)=m_0 \text{exp}\left(-x^2/d^2\right)$, where the magnetization in the bulk can take values $-1<m_0<1$ depending on the temperature, size, material details, etc. Under the reasonable assumption that $k_F^{-1}\ll d \ll \xi_\text{1D}$, with $\xi_\text{1D}$ the coherence length in the 1DNW, the FM contact becomes a ``point-like'' perturbation from the perspective of the chiral fields $\psi_{\nu\sigma}$, i.e., $m_z\left(x\right)\sim m_0 d \sqrt{\pi} \delta\left(x\right)$. Therefore, the Hamiltonian writes  $H^\text{1D}_{M} \simeq \frac{V m_0 d \sqrt{\pi}}{2}\sum_{r=L,R}\left[\psi_{r\uparrow}^{\dagger}\left(0\right)\psi_{r\uparrow}\left(0\right)-\psi_{r\downarrow}^{\dagger}\left(0\right)\psi_{r\downarrow}\left(0\right)\right]$, where we have neglected oscillatory terms  $\sim e^{\pm i 2k_Fx}$). In fermion language, this effective delta-like potential can be eliminated  with a gauge transformation $\psi_{r\sigma}\rightarrow \tilde{\psi}_{r\sigma}=e^{i2r\sigma\delta_0\Theta\left(x\right)}\psi_{r\sigma}$, where we have defined the phase shift $\delta_0= \arctan\left(\frac{V m_0 d\sqrt{\pi}}{4 v_F}\right)$, and where $\Theta\left(x\right)$ is the unit step function. The bosonized Hamiltonian of the FM contact then writes $H^\text{1D}_{M} = -2 v_F \delta_0 \frac{\partial_{x}\phi_{s}\left(0\right)}{\pi}$. Experimental values $k_F^{-1}\sim 20$ nm~\cite{Jespersen2009} and $\xi_\text{1D}\approx 260$ nm~\cite{Albrecht2016} are encouraging for the realization of this device. Note that in the absence of pairing the Hamiltonian is akin to the X-ray edge problem~\cite{Gogolin1988,Giamarchi2004}.

An interesting feature of the bosonic representation is that it allows to obtain the total spin and parity of the ground state from simple expressions
\begin{align}
s_{z}&=-\int_{-L_\text{W}}^{L_\text{W}}dx\,\frac{\partial_{x}\phi_{s}\left(x\right)}{2\pi}=-\frac{\Delta \phi_{s}}{2\pi},\label{eq:sz}\\
J&=\int_{-L_\text{W}}^{L_\text{W}}dx\,\frac{\partial_{x}\theta_{c}\left(x\right)}{\pi}=\frac{\Delta\theta_c}{\pi}.\label{eq:J}
\end{align}
The physical significance of $J$ can be clarified imposing periodic boundary conditions in the problem, i.e., $\psi_{r\sigma}(-L_\text{W})=\psi_{r\sigma}(L_\text{W})$. In this case, $J$ can only take integer values which are related to the total number of particles $N$ through the relation $J+N \mod 2=0$. Then, since  $P=\left(-1\right)^N=\left(-1\right)^J$, we conclude that $J$ determines the fermion-parity of the ground state~\cite{Haldane1981}.

Note that in the terms $H_\Delta$ and $H_M^\text{1D}$, only the commuting fields $\phi_s$ and $\theta_c$ appear. This indicates that the Hamiltonian has a well-defined ``classical'' limit when $\Delta$ flows to strong coupling, i.e., when it becomes of the order of the ultraviolet cutoff $\Delta\left(\ell^*\right)\simeq \Lambda$. This classical limit is more conventiently captured defining the fields $\phi_\pm=\left(\phi_s\pm \theta_c\right)/2$, and their canonical conjugates $\partial_x\theta_\pm = \left(\partial_x\theta_s\pm\partial_x\phi_c\right)/2$, in terms of which we have
\begin{align}
H_0=&\int_{-L_W}^{L_W} dx \left\{\frac{K_\theta v_\theta}{4\pi}\left[\sum_{\mu=\pm}\left(\partial_x \theta_\mu\right)^2+ 2\zeta_\theta \partial_x\theta_+ \partial_x \theta_-\right]\right.\nonumber\\
+&\left.\frac{K_\phi v_\phi}{4\pi}\left[\sum_{\mu=\pm}\left(\partial_x \phi_\mu\right)^2 + 2\zeta_\phi \partial_x\phi_+ \partial_x \phi_-\right]\right\},\label{eq:H0}\\
H_\Delta=&\frac{\Delta}{\pi a}\int_{-L_W}^{L_W} dx\, \left[\cos{2\phi_+}+\cos{2\phi_-}\right],\label{eq:H_Delta}\\
H^\text{1D}_M=&-\frac{2v_F\delta_0}{\pi} \left[\partial_x \phi_+\left(0\right)+\partial_x \phi_-\left(0\right)\right],\label{eq:H_M1D}
\end{align}
where we have defined $K_\theta v_\theta=\left(v_c/K_c+v_F\right)$, $K_\phi v_\phi=\left(v_F+K_c v_c\right)$, and the helicity parameters $\zeta_\theta\equiv \left(v_c/K_c-v_F \right)/\left(v_c/K_c+v_F\right)$ and $\zeta_\phi\equiv \left(v_F-K_c v_c \right)/\left(v_F+K_c v_c\right)$.  Even though the classical limit of $H$ can be studied in the generic case, here we concentrate on the simpler but more insightful  point $\zeta_\phi=0$ (i.e., $K_c v_c=v_F$), where the classical static configurations of $\phi_\pm$ decouple. The generic case will be discussed elsewhere  \cite{Bortolin2019}. At the point $\zeta_\phi=0$, we obtain the decoupled equations
\begin{equation}
\partial^2_x\phi_\mu=-g\sin2\phi_\mu+2\delta_0\delta^\prime(x).\quad \left(\mu=\pm\right)\label{eq:eq_phi_pm}
\end{equation}
Here $g=2\Delta/av_F=\xi^{-2}_\text{1D}$, and  $\delta^\prime(x)$ is the derivative of the Dirac delta-function, which has to be interpreted in terms of the approximant $\delta\left(x\right)=\lim_{a\rightarrow 0} f\left(x,a\right)$ as $\delta^\prime(x)=\lim_{a\rightarrow 0} \partial_x f\left(x,a\right)$.

\emph{3D Case}.---It is illuminating to briefly turn to the well-known case of a magnetic impurity in a 3D superconductor~\cite{Yu1965,Shiba1968,Rusinov1968}. For a spherically-symmetric problem, only the $s$-wave component of the 3D Fermi field 
$\Psi^\text{s}_{\sigma}\left(\mathbf{r}\right)$ couples to a point-like impurity placed at the origin. Then, linearizing the spectrum around the Fermi energy, we can write 
$\Psi^\text{s}_{\sigma}\left(\mathbf{r}\right) =\frac{1}{i2\sqrt{\pi}r}\left[e^{ik_{F}r}\psi_{L\sigma}\left(r\right)-e^{-ik_{F}r}\psi_{R\sigma}\left(r\right)\right]$, where $0\leq r\leq \infty$ is the radial coordinate. Equivalently, the half-line can be unfolded to the whole line by defining
\begin{equation}\label{eq:boundary_condition}
\psi_{r,\sigma}\left(-x\right)\equiv\psi_{-r,\sigma}\left(x\right),\quad \left(r=L,R\right)
\end{equation}
and keeping a \emph{single chirality} for each fermion. This definition includes the boundary condition  $\psi_{R\sigma}(0)=\psi_{L\sigma}(0)$  at $x=0$, which guarantees that $\Psi^\text{s}_\sigma\left(x\right)$ is finite at the origin. This procedure is standard and we refer the reader to Refs. \onlinecite{Giamarchi2004, Gogolin1988} for details. In what follows, when treating the 3D case, we will keep only the branches $\psi_{L\uparrow}$ and $\psi_{R\downarrow}$. Imposing a hard-wall boundary condition in an sphere of radius $R$ (i.e., $\Psi^\text{s}_\sigma(R)=0$), induces periodic boundary conditions on the effective 1D chiral fermions, $\psi_{r\sigma}(-R)=\psi_{r\sigma}(R)$. Therefore, we can write
\begin{equation}
H_0^\mathrm{3D}=-iv_F\int_{-R}^R dx\,\left[\psi^\dagger_{R\downarrow}\partial_x\psi_{R\downarrow} -\psi^\dagger_{L\uparrow}\partial_x\psi_{L\uparrow} \right],\label{eq:SHI}
\end{equation}
which upon bosonization results $H^\text{3D}_{0}=\frac{v_F}{2\pi}\int_{-R}^R  dx\,\left[(\partial_{x}\phi_{-})^{2}+(\partial_{x}\theta_{-})^{2}\right]$, where \emph{only the field $\phi_-$ appears}. Interestingly, Eq. (\ref{eq:SHI}) is the same Hamiltonian that describes the edge states of a spin-Hall insulator \cite{Wu06_Helical_liquid_and_the_QSHE}.  In addition, since the (repulsive) interactions are efficiently screened in a 3D superconductor, we have $K_c=K_s=1$ and $v_s=v_c=v_F$, and the system is automatically at the  point $\zeta_\phi=\zeta_\theta=0$.  On the other hand, it is easy to see that  due to the elimination of the chiral fields $\psi_{R\uparrow}, \psi_{L\downarrow}$, the pairing Hamiltonian in the $s$-wave channel writes $H^\text{3D}_\Delta=\Delta \int_{-R}^{R}  dx\, \left[\psi^\dagger_{L\uparrow}\psi^\dagger_{R\downarrow}+\hc\right]=\frac{\Delta}{\pi a} \int_{-R}^R  dx\, \cos{2\phi_{-}}$, where here $\Delta$ is an intrinsic interaction. Finally, the presence of a classical magnetic impurity is usually modelled using the $s$-$d$ Hamiltonian~\cite{Yu1965,Shiba1968,Rusinov1968}: $H^\text{3D}_{M} =\frac{J S^{z}}{2}\left[\Psi^{\text{s}\dagger}_{\uparrow}\left(0\right)\Psi^\text{s}_{\uparrow}\left(0\right)-\Psi^{\text{s}\dagger}_{\downarrow}\left(0\right)\Psi^\text{s}_{\downarrow}\left(0\right)\right]$, where $J$ is the exchange coupling and $S^{z}=\pm S$ is the $z$-component of the magnetic impurity $S$, assumed as an Ising spin. In terms of the chiral  fields, this term writes $H^\text{3D}_{M} =\frac{J k^2_F S^{z}}{2\pi }\left[\psi_{L\uparrow}^{\dagger}\left(0\right)\psi_{L\uparrow}\left(0\right)-\psi_{R\downarrow}^{\dagger}\left(0\right)\psi_{R\downarrow}\left(0\right)\right]$, where we have used Eq. (\ref{eq:boundary_condition}) and the result $\lim_{r\rightarrow 0}\Psi^\text{s}_{\sigma}\left(\mathbf{r}\right)=\frac{k_F}{2\sqrt{\pi}}\left[\psi_{L\sigma}\left(0\right)+\psi_{R\sigma}\left(0\right)\right]$. Again, $H^\text{3D}_{M}$ can be absorbed via the gauge transformation $\psi_{L\uparrow(R\downarrow)}\rightarrow \tilde{\psi}_{L\uparrow(R\downarrow)}=e^{\mp i2\delta_0 \Theta\left(x\right)}\psi_{L\uparrow(R\downarrow)}$, where $\delta_0=\arctan{\left[J  S^z \rho_0 \pi/2\right]}$ is the $s$-wave phase shift (note that we have used the definition of the  density of states at the Fermi level $\rho_0=k^2_F/4\pi v_F$)\cite{Balatsky2006}. Then, in bosonic language we obtain the expression $H^\text{3D}_{M} =-2v_F\delta_0 \partial_x \phi_-\left(0\right)/\pi$. In this way, the connection between $H^\text{1D}$  and  $H^\text{3D}$ can be clearly seen: the 3D Hamiltonian is identical to the 1D Hamiltonian at the  point $\zeta_\phi=0$ with \emph{half the degrees of freedom} (i.e., the field $\phi_+$ is absent, and the system becomes a \emph{helical liquid} \cite{Wu06_Helical_liquid_and_the_QSHE}. Consequently, the classical configuration of the field $\phi_-$ is also given by Eq. (\ref{eq:eq_phi_pm}).

\emph{Results}---We now turn to the resolution of Eq. (\ref{eq:eq_phi_pm}). To that end, we write the fields $\phi_\pm$ as
\begin{equation}\label{eq:split}
\phi_\mu(x)=\eta_\mu\frac{\pi}{2}+2\delta_0 \Theta(x)+\varphi_\mu(x)\quad \left(\mu=\pm\right),
\end{equation}
which allows to eliminate the term $\delta^\prime\left(x\right)$. Here $\varphi_\mu(x)$ is a continuous and smooth function at $x=0$. We have found that the soliton/antisoliton-like  functions ($\eta_\mu=\pm1$)
\begin{equation}
\varphi_\mu=\begin{cases}
2\arctan e^{\eta_\mu \sqrt{2g}\left(x-x_{0}\right)}&\text{for }x<0,\\
2\arctan e^{\eta_\mu \sqrt{2g}\left(x+x_{0}\right)}+\pi\eta_\mu l_\mu-2 \delta_0& \text{for }x\geqslant0.
\end{cases}\label{eq:varphi_s_helical}
\end{equation}
are exact solutions of Eq. (\ref{eq:eq_phi_pm}). Here, $l_\mu$ is an integer which determine the different minima of the $\cos 2\phi_\mu$ potential, and $x_{0}$ is a parameter which must be chosen so that the continuity condition at the origin $\varphi_\mu(0^{-})=\varphi_\mu(0^{+})$ is verified. From this condition we obtain the equation
\begin{align}\label{eq:x0}
\arctan\left[\sinh\left({\eta_\mu \sqrt{2g}x_{0}}\right)\right]& =\delta_{0}-\eta_\mu l_\mu\frac{\pi}{2},
\end{align}
from where $x_0\equiv x_0\left(\delta_0,\eta_\mu,l_\mu\right)$ is obtained. Once the value of $\delta_0$ and the nature of the solution (i.e., soliton or anti-soliton $\eta_\mu=\pm 1$) are fixed, the integer $l_\mu$ becomes restricted to only two possibilities since the left hand side of Eq. (\ref{eq:x0}) can only vary between $-\pi/2$ and $\pi/2$.  

\begin{figure}
  \centering
	\includegraphics[viewport=30 0 400 410,clip,width=\columnwidth]{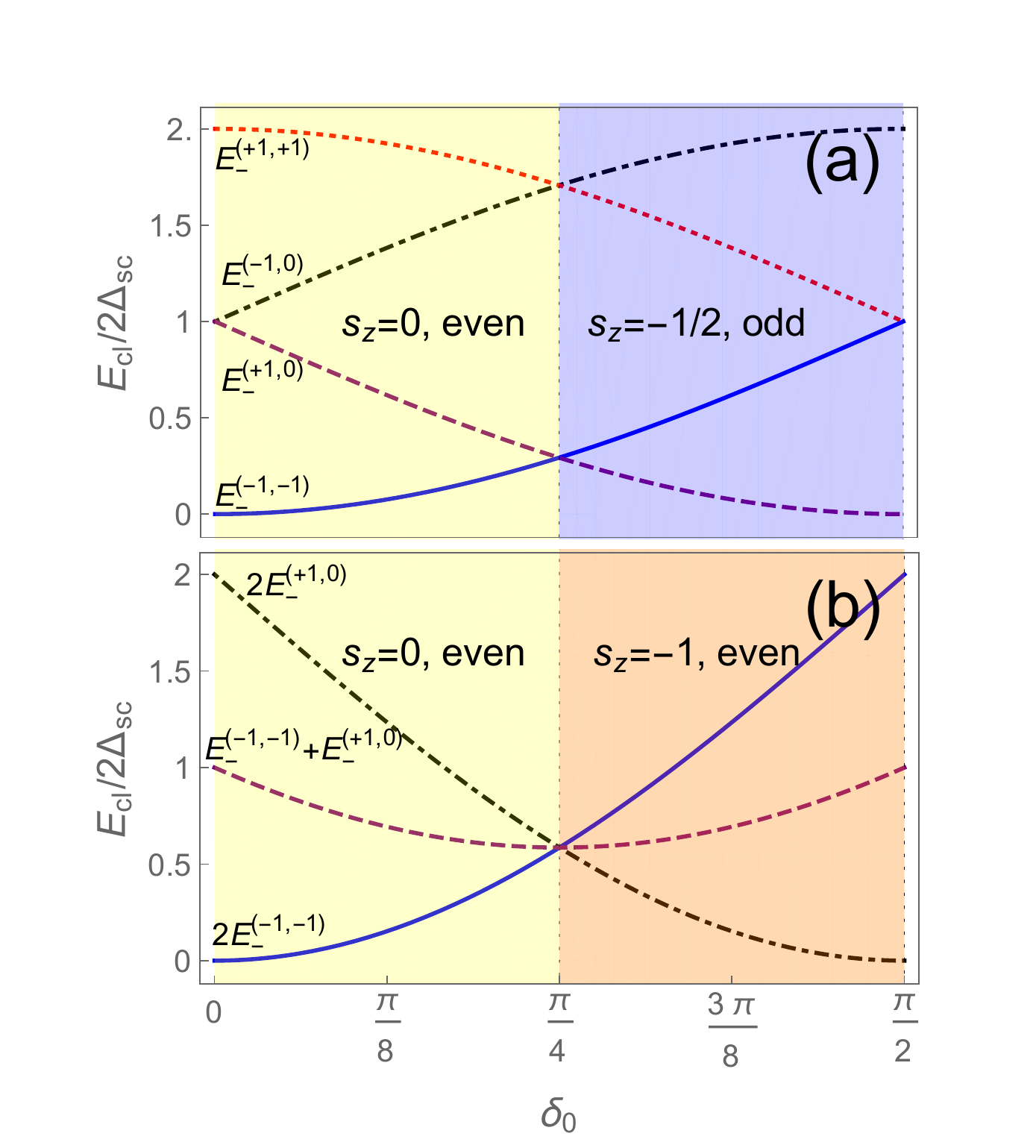}
  \caption{Phase diagram of the Shiba impurity at $\zeta_\phi=0$ where $v_c K_c=v_F$. 
 a) 3D Case: Energy of the different branches as a function of the phase shift $\delta_0$ (see Eq. (\ref{eq:energy_soliton_shiba})). At the critical value $\delta_0=\pi/4$ the system exhibits a phase transition from an even-parity singlet state to an odd-parity spin-1/2 doublet. b) 1D Case: Idem for the case of a 1DNW, where both fields $\phi_\pm$ are present. At $\delta_0=\pi/4$, a 0-0 transition occurs  from an even-parity singlet state to an even-parity $s_z=-1$ state. }\label{fig:phases}
\end{figure}

Of special importance here is the physical meaning associated to the topological charge of the solitonic solutions (\ref{eq:split}) and (\ref{eq:varphi_s_helical}). Replacing them into Eqs. (\ref{eq:sz}) and (\ref{eq:J}), we obtain $s_z=-\frac{1}{2} \sum_{\mu=\pm} \eta_\mu\left(l_\mu +1\right)$ and $J=\sum_{\mu=\pm} \left(\mu\right)\eta_\mu\left(l_\mu +1\right)$. In addition,  replacing them into the classical part of the Hamiltonian (\ref{eq:H0})-(\ref{eq:H_M1D}) allows to obtain 
their associated energy, i.e., 
$E\left(\delta_0\right)=\sum_{\mu=\pm}
E^{\left(\eta_\mu,l_\mu\right)}_\mu\left(\delta_0\right)$, where
$E^{\left(\eta_\mu,l_\mu\right)}_\mu\left(\delta_0\right)=\frac{v_F}{2\pi}\int^{L_W}_{-L_W} dx \left[\left(\partial_x \phi_\mu  -2\delta_0 \delta\left(x\right)\right)^2 + g \left(\cos 2\phi_\mu +1\right)\right]$, where we have substracted a constant background term so that the ground-state energy for $\delta_0=0$ (i.e., decoupled impurity) is zero. The numbers $\eta_\mu$ and $l_\mu$ determine a particular 
``energy branch'' $E^{\left(\eta_\mu,l_\mu\right)}_\mu\left(\delta_0\right)$ associated to that solution (see Fig. \ref{fig:phases}). In the limit $L_W\rightarrow \infty$ (or $R\rightarrow \infty$) we obtain the analytical result
\begin{align}
\frac{E^{\left(\eta_\mu,l_\mu\right)}_\mu\left(\delta_0\right)}{2\Delta_{\text{sc}}}&=1-\text{sign}\left(x_0\right)\left|\sin\left(\delta_0-\frac{\pi \eta_\mu  l_\mu}{2}\right)\right|,\label{eq:energy_soliton_shiba}
\end{align}
where we have defined $\Delta_{\text{sc}}\equiv\frac{v_{F}\sqrt{2g}}{\pi}=\frac{1}{\pi}\sqrt{\frac{2\Delta v_{F}}{a}}$. 

It is instructive to analyze first  the results for a Shiba impurity in a 3D host. For a decoupled impurity (i.e., when $\delta_0= 0$), the ground-state configuration of the superconductor is described by a constant $\phi_-= \frac{\pi}{2}$, which trivially minimizes the pairing term $\cos 2\phi_-$.  This situation physically corresponds to the ``classical'' BCS ground state. Our solutions (\ref{eq:split}) and (\ref{eq:varphi_s_helical}) precisely recover this behavior for $\left(\eta_-,l_-\right)$=$(-1, -1)$, for which Eq. (\ref{eq:x0})  yields  $x_0\rightarrow \infty$,  and Eq. (\ref{eq:energy_soliton_shiba}) yields the energy $E_{-}^{\left(-1,-1\right)}\left(0\right)=0$ (see continous blue line in Fig. \ref{fig:phases}(a)). This ground state has topological numbers $s_z=0$ and $J=0$ (i.e., even-parity singlet), as expected physically for a BCS superconductor. On the other hand, the solution with $\eta_-=+1$  ($\eta_-=-1$) and $l_-=0$ corresponds to a topological kink (anti-kink) connecting the minima $\phi_-= \frac{\pi}{2}$ at $x=-\infty$ and $\phi_-= \frac{3\pi}{2}$  ($\phi_-= -\frac{\pi}{2}$) at $x=\infty$, and corresponds to an excited state with $s_z=-1/2$ ($s_z=1/2$) and $J=-1$ ($J=1$) (odd-parity doublet). We interpret this state as having an extra (one less) fermion with respect to the BCS ground state and has energy $E_{-}^{\left(\pm 1,0\right)}\left(0\right)=2\Delta_\text{sc}$ (dashed purple line in Fig. \ref{fig:phases}(a)). Other excited branches are shown in that figure. As the value of $\delta_0$ is increased, the branches $E_{-}^{\left(\pm -1,-1\right)}$ and $E_{-}^{\left(\pm 1,0\right)}$ approach to each other and eventually cross at the value $\delta^\text{c}_0=\pi/4$, signalling a quantum phase  transition (i.e., the $0-\pi$ transition) at which the ground state abruptly changes spin and parity. Interestingly, our formalism allows to \emph{exactly} recover the critical value of the transition obtained from the crossing of Shiba levels $E_\text{Shiba}=\pm \frac{1-\alpha^2}{1+\alpha^2}$ (where $\alpha=J\rho_0 \pi S^z/2$)~\cite{Shiba1968,Balatsky2006} (note that precisely when $E_\text{Shiba}=0$, we obtain $\delta^\text{c}_0=\pi/4$). 

In the case of the 1DNW, one can immediately determine the quantum phase diagram exploiting the degeneracy of the solutions $\phi_\pm$ at $\zeta_\phi=0$. The ground-state energy is given by the sum of the lowest degenerate branches at a particular value of $\delta_0$. For $\delta_0<\pi/4$, the ground-state energy is given by $\sum_\mu E^{\left(-1,-1\right)}_\mu \left(\delta_0\right)$ and corresponds to an even-parity singlet phase  with topological numbers $s_z=0$ and $J=0$ (see blue line in Fig. \ref{fig:phases}(b)). On the other hand, for $\delta_0>\pi/4$ the ground state is magnetic (topological number $s_z=-1$) and has even parity ($J=2$). Therefore, in this case the value $\delta^\text{c}_0=\pi/4$  indicates a ``0-0 transition'' that \emph{preserves parity but not the spin}, in stark contrast to the 3D case. This result can be understood due to the presence of \emph{two} independent bosonic modes  $\phi_\pm$ in a 1D geometry. Physically, each one of these modes are related to the fermionic bilinears $\cos{2\phi_+} \sim \psi^\dagger_{R\uparrow}\psi^\dagger_{L\downarrow}$ and $\cos{2\phi_-}\sim \psi^\dagger_{L\uparrow}\psi^\dagger_{R\downarrow}$, each one capable of binding one Shiba state with $s_z=-1/2$ at the critical point. Therefore, in the 1D geometry the transition occurs when these two independent Shiba states simultaneously cross the Fermi energy and become occupied. Thus, the spin quantum number of the ground state jumps from $s_z=0$ to $s_z=-1$, whereas the parity remains unchanged. Experimentally, this effect would imply the absence of sign-reversal of a supercurrent flowing through the 1DNW.

Note that for this effect to occur it is crucial to avoid the mixing of the  fields $\phi_\pm$. To that end, backwards scattering terms $\sim e^{i2k_\text{F}x}e^{i\phi_c\left(x\right)}$ in the Hamiltonian (\ref{eq:H_M1D}) should be absent.  Although in principle backscattering terms are present in the expression of the spin density $\rho_s\left(x\right)$, due the assumption of a mesoscopic FM contact of width $d\gg k^{-1}_\text{F}$ this mechanism can be strongly suppressed. Note that this would not be the case in the more usual superconductor-quantum dot-superconductor geometry. Another requirement for the observation of the 0-0 transition is that the system obeys $K_c v_c=v_F$. Away from this point the fields $\phi_\pm$ become mixed, enabling the study of potentially interesting phenomena within the theoretical framework of bosonization.

\acknowledgements{
This work is partially supported by ANPCyT,  Argentina. AML acknowledges the use of Toko Cluster facilities from FCEN-UNCuyo, which is part of the SNCAD-MinCyT, Argentina.}


\begin{thebibliography}{36}%
\makeatletter
\providecommand \@ifxundefined [1]{%
 \@ifx{#1\undefined}
}%
\providecommand \@ifnum [1]{%
 \ifnum #1\expandafter \@firstoftwo
 \else \expandafter \@secondoftwo
 \fi
}%
\providecommand \@ifx [1]{%
 \ifx #1\expandafter \@firstoftwo
 \else \expandafter \@secondoftwo
 \fi
}%
\providecommand \natexlab [1]{#1}%
\providecommand \enquote  [1]{``#1''}%
\providecommand \bibnamefont  [1]{#1}%
\providecommand \bibfnamefont [1]{#1}%
\providecommand \citenamefont [1]{#1}%
\providecommand \href@noop [0]{\@secondoftwo}%
\providecommand \href [0]{\begingroup \@sanitize@url \@href}%
\providecommand \@href[1]{\@@startlink{#1}\@@href}%
\providecommand \@@href[1]{\endgroup#1\@@endlink}%
\providecommand \@sanitize@url [0]{\catcode `\\12\catcode `\$12\catcode
  `\&12\catcode `\#12\catcode `\^12\catcode `\_12\catcode `\%12\relax}%
\providecommand \@@startlink[1]{}%
\providecommand \@@endlink[0]{}%
\providecommand \url  [0]{\begingroup\@sanitize@url \@url }%
\providecommand \@url [1]{\endgroup\@href {#1}{\urlprefix }}%
\providecommand \urlprefix  [0]{URL }%
\providecommand \Eprint [0]{\href }%
\providecommand \doibase [0]{http://dx.doi.org/}%
\providecommand \selectlanguage [0]{\@gobble}%
\providecommand \bibinfo  [0]{\@secondoftwo}%
\providecommand \bibfield  [0]{\@secondoftwo}%
\providecommand \translation [1]{[#1]}%
\providecommand \BibitemOpen [0]{}%
\providecommand \bibitemStop [0]{}%
\providecommand \bibitemNoStop [0]{.\EOS\space}%
\providecommand \EOS [0]{\spacefactor3000\relax}%
\providecommand \BibitemShut  [1]{\csname bibitem#1\endcsname}%
\let\auto@bib@innerbib\@empty
\bibitem [{\citenamefont {Yu}(1965)}]{Yu1965}%
  \BibitemOpen
  \bibfield  {author} {\bibinfo {author} {\bibfnamefont {L.}~\bibnamefont
  {Yu}},\ }\href {\doibase 10.7498/aps.21.75} {\bibfield  {journal} {\bibinfo
  {journal} {Acta Phys. Sin.}\ }\textbf {\bibinfo {volume} {21}},\ \bibinfo
  {pages} {75} (\bibinfo {year} {1965})}\BibitemShut {NoStop}%
\bibitem [{\citenamefont {Shiba}(1968)}]{Shiba1968}%
  \BibitemOpen
  \bibfield  {author} {\bibinfo {author} {\bibfnamefont {H.}~\bibnamefont
  {Shiba}},\ }\href@noop {} {\bibfield  {journal} {\bibinfo  {journal} {Prog.
  Theor. Phys.}\ }\textbf {\bibinfo {volume} {40}},\ \bibinfo {pages} {435}
  (\bibinfo {year} {1968})}\BibitemShut {NoStop}%
\bibitem [{\citenamefont {Rusinov}(1968)}]{Rusinov1968}%
  \BibitemOpen
  \bibfield  {author} {\bibinfo {author} {\bibfnamefont {A.~I.}\ \bibnamefont
  {Rusinov}},\ }\href {\doibase
  http://www.jetpletters.ac.ru/ps/1658/article_25295.shtml} {\bibfield
  {journal} {\bibinfo  {journal} {Zh. Eksp. Teor. Fiz. Pisma. Red.}\ }\textbf
  {\bibinfo {volume} {9}},\ \bibinfo {pages} {146} (\bibinfo {year} {1968})},\
  \bibinfo {note} {[JETP Lett. 9, 85 (1969)]}\BibitemShut {NoStop}%
\bibitem [{\citenamefont {Balatsky}\ \emph {et~al.}(2006)\citenamefont
  {Balatsky}, \citenamefont {Vekhter},\ and\ \citenamefont
  {Zhu}}]{Balatsky2006}%
  \BibitemOpen
  \bibfield  {author} {\bibinfo {author} {\bibfnamefont {A.~V.}\ \bibnamefont
  {Balatsky}}, \bibinfo {author} {\bibfnamefont {I.}~\bibnamefont {Vekhter}}, \
  and\ \bibinfo {author} {\bibfnamefont {J.-X.}\ \bibnamefont {Zhu}},\ }\href
  {\doibase 10.1103/revmodphys.78.373} {\bibfield  {journal} {\bibinfo
  {journal} {Rev. Mod. Phys.}\ }\textbf {\bibinfo {volume} {78}},\ \bibinfo
  {pages} {373} (\bibinfo {year} {2006})}\BibitemShut {NoStop}%
\bibitem [{\citenamefont {De~Franceschi}\ \emph {et~al.}(2010)\citenamefont
  {De~Franceschi}, \citenamefont {Kouwenhoven}, \citenamefont
  {Sch{\"o}nenberger},\ and\ \citenamefont {Wernsdorfer}}]{DeFranceschi2010}%
  \BibitemOpen
  \bibfield  {author} {\bibinfo {author} {\bibfnamefont {S.}~\bibnamefont
  {De~Franceschi}}, \bibinfo {author} {\bibfnamefont {L.}~\bibnamefont
  {Kouwenhoven}}, \bibinfo {author} {\bibfnamefont {C.}~\bibnamefont
  {Sch{\"o}nenberger}}, \ and\ \bibinfo {author} {\bibfnamefont
  {W.}~\bibnamefont {Wernsdorfer}},\ }\href {\doibase 10.1038/nnano.2010.173}
  {\bibfield  {journal} {\bibinfo  {journal} {Nature Nanotechnology}\ }\textbf
  {\bibinfo {volume} {5}},\ \bibinfo {pages} {703} (\bibinfo {year}
  {2010})}\BibitemShut {NoStop}%
\bibitem [{\citenamefont {Deacon}\ \emph {et~al.}(2010)\citenamefont {Deacon},
  \citenamefont {Tanaka}, \citenamefont {Oiwa}, \citenamefont {Sakano},
  \citenamefont {Yoshida}, \citenamefont {Shibata}, \citenamefont {Hirakawa},\
  and\ \citenamefont {Tarucha}}]{Deacon2010}%
  \BibitemOpen
  \bibfield  {author} {\bibinfo {author} {\bibfnamefont {R.~S.}\ \bibnamefont
  {Deacon}}, \bibinfo {author} {\bibfnamefont {Y.}~\bibnamefont {Tanaka}},
  \bibinfo {author} {\bibfnamefont {A.}~\bibnamefont {Oiwa}}, \bibinfo {author}
  {\bibfnamefont {R.}~\bibnamefont {Sakano}}, \bibinfo {author} {\bibfnamefont
  {K.}~\bibnamefont {Yoshida}}, \bibinfo {author} {\bibfnamefont
  {K.}~\bibnamefont {Shibata}}, \bibinfo {author} {\bibfnamefont
  {K.}~\bibnamefont {Hirakawa}}, \ and\ \bibinfo {author} {\bibfnamefont
  {S.}~\bibnamefont {Tarucha}},\ }\href {\doibase
  10.1103/PhysRevLett.104.076805} {\bibfield  {journal} {\bibinfo  {journal}
  {Phys. Rev. Lett.}\ }\textbf {\bibinfo {volume} {104}},\ \bibinfo {pages}
  {076805} (\bibinfo {year} {2010})}\BibitemShut {NoStop}%
\bibitem [{\citenamefont {Mart\'{\i}n-Rodero}\ and\ \citenamefont
  {Yeyati}(2011)}]{MartinRodero2011}%
  \BibitemOpen
  \bibfield  {author} {\bibinfo {author} {\bibfnamefont {A.}~\bibnamefont
  {Mart\'{\i}n-Rodero}}\ and\ \bibinfo {author} {\bibfnamefont {A.~L.}\
  \bibnamefont {Yeyati}},\ }\href {\doibase 10.1080/00018732.2011.624266}
  {\bibfield  {journal} {\bibinfo  {journal} {Advances in Physics}\ }\textbf
  {\bibinfo {volume} {60}},\ \bibinfo {pages} {899} (\bibinfo {year}
  {2011})}\BibitemShut {NoStop}%
\bibitem [{\citenamefont {Maurand}\ \emph {et~al.}(2012)\citenamefont
  {Maurand}, \citenamefont {Meng}, \citenamefont {Bonet}, \citenamefont
  {Florens}, \citenamefont {Marty},\ and\ \citenamefont
  {Wernsdorfer}}]{Maurand2012}%
  \BibitemOpen
  \bibfield  {author} {\bibinfo {author} {\bibfnamefont {R.}~\bibnamefont
  {Maurand}}, \bibinfo {author} {\bibfnamefont {T.}~\bibnamefont {Meng}},
  \bibinfo {author} {\bibfnamefont {E.}~\bibnamefont {Bonet}}, \bibinfo
  {author} {\bibfnamefont {S.}~\bibnamefont {Florens}}, \bibinfo {author}
  {\bibfnamefont {L.}~\bibnamefont {Marty}}, \ and\ \bibinfo {author}
  {\bibfnamefont {W.}~\bibnamefont {Wernsdorfer}},\ }\href {\doibase
  10.1103/PhysRevX.2.011009} {\bibfield  {journal} {\bibinfo  {journal} {Phys.
  Rev. X}\ }\textbf {\bibinfo {volume} {2}},\ \bibinfo {pages} {011009}
  (\bibinfo {year} {2012})}\BibitemShut {NoStop}%
\bibitem [{\citenamefont {Nayak}\ \emph {et~al.}(2008)\citenamefont {Nayak},
  \citenamefont {Simon}, \citenamefont {Stern}, \citenamefont {Freedman},\ and\
  \citenamefont {{Das Sarma}}}]{Nayak2008}%
  \BibitemOpen
  \bibfield  {author} {\bibinfo {author} {\bibfnamefont {C.}~\bibnamefont
  {Nayak}}, \bibinfo {author} {\bibfnamefont {S.~H.}\ \bibnamefont {Simon}},
  \bibinfo {author} {\bibfnamefont {A.}~\bibnamefont {Stern}}, \bibinfo
  {author} {\bibfnamefont {M.}~\bibnamefont {Freedman}}, \ and\ \bibinfo
  {author} {\bibfnamefont {S.}~\bibnamefont {{Das Sarma}}},\ }\href {\doibase
  10.1103/RevModPhys.80.1083} {\bibfield  {journal} {\bibinfo  {journal} {Rev.
  Mod. Phys.}\ }\textbf {\bibinfo {volume} {80}},\ \bibinfo {pages} {1083}
  (\bibinfo {year} {2008})}\BibitemShut {NoStop}%
\bibitem [{\citenamefont {Nadj-Perge}\ \emph {et~al.}(2013)\citenamefont
  {Nadj-Perge}, \citenamefont {Drozdov}, \citenamefont {Bernevig},\ and\
  \citenamefont {Yazdani}}]{Nadj-Perge2013}%
  \BibitemOpen
  \bibfield  {author} {\bibinfo {author} {\bibfnamefont {S.}~\bibnamefont
  {Nadj-Perge}}, \bibinfo {author} {\bibfnamefont {I.~K.}\ \bibnamefont
  {Drozdov}}, \bibinfo {author} {\bibfnamefont {B.~A.}\ \bibnamefont
  {Bernevig}}, \ and\ \bibinfo {author} {\bibfnamefont {A.}~\bibnamefont
  {Yazdani}},\ }\href {\doibase 10.1103/physrevb.88.020407} {\bibfield
  {journal} {\bibinfo  {journal} {Phys. Rev. B}\ }\textbf {\bibinfo {volume}
  {88}},\ \bibinfo {pages} {020407} (\bibinfo {year} {2013})}\BibitemShut
  {NoStop}%
\bibitem [{\citenamefont {Klinovaja}\ \emph {et~al.}(2013)\citenamefont
  {Klinovaja}, \citenamefont {Stano}, \citenamefont {Yazdani},\ and\
  \citenamefont {Loss}}]{Klinovaja2013}%
  \BibitemOpen
  \bibfield  {author} {\bibinfo {author} {\bibfnamefont {J.}~\bibnamefont
  {Klinovaja}}, \bibinfo {author} {\bibfnamefont {P.}~\bibnamefont {Stano}},
  \bibinfo {author} {\bibfnamefont {A.}~\bibnamefont {Yazdani}}, \ and\
  \bibinfo {author} {\bibfnamefont {D.}~\bibnamefont {Loss}},\ }\href {\doibase
  10.1103/PhysRevLett.111.186805} {\bibfield  {journal} {\bibinfo  {journal}
  {Phys. Rev. Lett.}\ }\textbf {\bibinfo {volume} {111}},\ \bibinfo {pages}
  {186805} (\bibinfo {year} {2013})}\BibitemShut {NoStop}%
\bibitem [{\citenamefont {Braunecker}\ and\ \citenamefont
  {Simon}(2013)}]{Braunecker2013}%
  \BibitemOpen
  \bibfield  {author} {\bibinfo {author} {\bibfnamefont {B.}~\bibnamefont
  {Braunecker}}\ and\ \bibinfo {author} {\bibfnamefont {P.}~\bibnamefont
  {Simon}},\ }\href {\doibase 10.1103/PhysRevLett.111.147202} {\bibfield
  {journal} {\bibinfo  {journal} {Phys. Rev. Lett.}\ }\textbf {\bibinfo
  {volume} {111}},\ \bibinfo {pages} {147202} (\bibinfo {year}
  {2013})}\BibitemShut {NoStop}%
\bibitem [{\citenamefont {Pientka}\ \emph {et~al.}(2013)\citenamefont
  {Pientka}, \citenamefont {Glazman},\ and\ \citenamefont {von
  Oppen}}]{Pientka2013}%
  \BibitemOpen
  \bibfield  {author} {\bibinfo {author} {\bibfnamefont {F.}~\bibnamefont
  {Pientka}}, \bibinfo {author} {\bibfnamefont {L.~I.}\ \bibnamefont
  {Glazman}}, \ and\ \bibinfo {author} {\bibfnamefont {F.}~\bibnamefont {von
  Oppen}},\ }\href {\doibase 10.1103/physrevb.88.155420} {\bibfield  {journal}
  {\bibinfo  {journal} {Phys. Rev. B}\ }\textbf {\bibinfo {volume} {88}},\
  \bibinfo {pages} {155420} (\bibinfo {year} {2013})}\BibitemShut {NoStop}%
\bibitem [{\citenamefont {Kitaev}(2001)}]{Kitaev2001}%
  \BibitemOpen
  \bibfield  {author} {\bibinfo {author} {\bibfnamefont {A.~Y.}\ \bibnamefont
  {Kitaev}},\ }\href {\doibase 10.1070/1063-7869/44/10S/S29} {\bibfield
  {journal} {\bibinfo  {journal} {Phys. Usp.}\ }\textbf {\bibinfo {volume}
  {44}},\ \bibinfo {pages} {131} (\bibinfo {year} {2001})}\BibitemShut
  {NoStop}%
\bibitem [{\citenamefont {Nadj-Perge}\ \emph {et~al.}(2014)\citenamefont
  {Nadj-Perge}, \citenamefont {Drozdov}, \citenamefont {Li}, \citenamefont
  {Chen}, \citenamefont {Jeon}, \citenamefont {Seo}, \citenamefont {MacDonald},
  \citenamefont {Bernevig},\ and\ \citenamefont {Yazdani}}]{Nadj-Perge2014}%
  \BibitemOpen
  \bibfield  {author} {\bibinfo {author} {\bibfnamefont {S.}~\bibnamefont
  {Nadj-Perge}}, \bibinfo {author} {\bibfnamefont {I.~K.}\ \bibnamefont
  {Drozdov}}, \bibinfo {author} {\bibfnamefont {J.}~\bibnamefont {Li}},
  \bibinfo {author} {\bibfnamefont {H.}~\bibnamefont {Chen}}, \bibinfo {author}
  {\bibfnamefont {S.}~\bibnamefont {Jeon}}, \bibinfo {author} {\bibfnamefont
  {J.}~\bibnamefont {Seo}}, \bibinfo {author} {\bibfnamefont {A.~H.}\
  \bibnamefont {MacDonald}}, \bibinfo {author} {\bibfnamefont {B.~A.}\
  \bibnamefont {Bernevig}}, \ and\ \bibinfo {author} {\bibfnamefont
  {A.}~\bibnamefont {Yazdani}},\ }\href {\doibase 10.1126/science.1259327}
  {\bibfield  {journal} {\bibinfo  {journal} {Science}\ }\textbf {\bibinfo
  {volume} {346}},\ \bibinfo {pages} {602} (\bibinfo {year}
  {2014})}\BibitemShut {NoStop}%
\bibitem [{\citenamefont {Pawlak}\ \emph {et~al.}(2016)\citenamefont {Pawlak},
  \citenamefont {Kisiel}, \citenamefont {Klinovaja}, \citenamefont {Meier},
  \citenamefont {Kawai}, \citenamefont {Glatzel}, \citenamefont {Loss},\ and\
  \citenamefont {Meyer}}]{Pawlak2016}%
  \BibitemOpen
  \bibfield  {author} {\bibinfo {author} {\bibfnamefont {R.}~\bibnamefont
  {Pawlak}}, \bibinfo {author} {\bibfnamefont {M.}~\bibnamefont {Kisiel}},
  \bibinfo {author} {\bibfnamefont {J.}~\bibnamefont {Klinovaja}}, \bibinfo
  {author} {\bibfnamefont {T.}~\bibnamefont {Meier}}, \bibinfo {author}
  {\bibfnamefont {S.}~\bibnamefont {Kawai}}, \bibinfo {author} {\bibfnamefont
  {T.}~\bibnamefont {Glatzel}}, \bibinfo {author} {\bibfnamefont
  {D.}~\bibnamefont {Loss}}, \ and\ \bibinfo {author} {\bibfnamefont
  {E.}~\bibnamefont {Meyer}},\ }\href {\doibase 10.1038/npjqi.2016.35}
  {\bibfield  {journal} {\bibinfo  {journal} {npj Quantum Information}\
  }\textbf {\bibinfo {volume} {2}},\ \bibinfo {pages} {16035} (\bibinfo {year}
  {2016})}\BibitemShut {NoStop}%
\bibitem [{\citenamefont {Ruby}\ \emph {et~al.}(2015)\citenamefont {Ruby},
  \citenamefont {Pientka}, \citenamefont {Peng}, \citenamefont {von Oppen},
  \citenamefont {Heinrich},\ and\ \citenamefont {Franke}}]{Ruby2015}%
  \BibitemOpen
  \bibfield  {author} {\bibinfo {author} {\bibfnamefont {M.}~\bibnamefont
  {Ruby}}, \bibinfo {author} {\bibfnamefont {F.}~\bibnamefont {Pientka}},
  \bibinfo {author} {\bibfnamefont {Y.}~\bibnamefont {Peng}}, \bibinfo {author}
  {\bibfnamefont {F.}~\bibnamefont {von Oppen}}, \bibinfo {author}
  {\bibfnamefont {B.~W.}\ \bibnamefont {Heinrich}}, \ and\ \bibinfo {author}
  {\bibfnamefont {K.~J.}\ \bibnamefont {Franke}},\ }\href {\doibase
  10.1103/physrevlett.115.197204} {\bibfield  {journal} {\bibinfo  {journal}
  {Phys. Rev. Lett.}\ }\textbf {\bibinfo {volume} {115}},\ \bibinfo {pages}
  {197204} (\bibinfo {year} {2015})}\BibitemShut {NoStop}%
\bibitem [{\citenamefont {Sakurai}(1970)}]{Sakurai1970}%
  \BibitemOpen
  \bibfield  {author} {\bibinfo {author} {\bibfnamefont {A.}~\bibnamefont
  {Sakurai}},\ }\href {\doibase 10.1143/ptp.44.1472} {\bibfield  {journal}
  {\bibinfo  {journal} {Prog. Theor. Phys.}\ }\textbf {\bibinfo {volume}
  {44}},\ \bibinfo {pages} {1472} (\bibinfo {year} {1970})}\BibitemShut
  {NoStop}%
\bibitem [{\citenamefont {Franke}\ \emph {et~al.}(2011)\citenamefont {Franke},
  \citenamefont {Schulze},\ and\ \citenamefont {Pascual}}]{Franke2011}%
  \BibitemOpen
  \bibfield  {author} {\bibinfo {author} {\bibfnamefont {K.~J.}\ \bibnamefont
  {Franke}}, \bibinfo {author} {\bibfnamefont {G.}~\bibnamefont {Schulze}}, \
  and\ \bibinfo {author} {\bibfnamefont {J.~I.}\ \bibnamefont {Pascual}},\
  }\href {\doibase 10.1126/science.1202204} {\bibfield  {journal} {\bibinfo
  {journal} {Science}\ }\textbf {\bibinfo {volume} {332}},\ \bibinfo {pages}
  {940} (\bibinfo {year} {2011})}\BibitemShut {NoStop}%
\bibitem [{\citenamefont {Bauer}\ \emph {et~al.}(2013)\citenamefont {Bauer},
  \citenamefont {Pascual},\ and\ \citenamefont {Franke}}]{Bauer2013}%
  \BibitemOpen
  \bibfield  {author} {\bibinfo {author} {\bibfnamefont {J.}~\bibnamefont
  {Bauer}}, \bibinfo {author} {\bibfnamefont {J.~I.}\ \bibnamefont {Pascual}},
  \ and\ \bibinfo {author} {\bibfnamefont {K.~J.}\ \bibnamefont {Franke}},\
  }\href {\doibase 10.1103/PhysRevB.87.075125} {\bibfield  {journal} {\bibinfo
  {journal} {Phys. Rev. B}\ }\textbf {\bibinfo {volume} {87}},\ \bibinfo
  {pages} {075125} (\bibinfo {year} {2013})}\BibitemShut {NoStop}%
\bibitem [{\citenamefont {Lee}\ \emph {et~al.}(2012)\citenamefont {Lee},
  \citenamefont {Jiang}, \citenamefont {Aguado}, \citenamefont {Katsaros},
  \citenamefont {Lieber},\ and\ \citenamefont {De~Franceschi}}]{Lee2012}%
  \BibitemOpen
  \bibfield  {author} {\bibinfo {author} {\bibfnamefont {E.~J.~H.}\
  \bibnamefont {Lee}}, \bibinfo {author} {\bibfnamefont {X.}~\bibnamefont
  {Jiang}}, \bibinfo {author} {\bibfnamefont {R.}~\bibnamefont {Aguado}},
  \bibinfo {author} {\bibfnamefont {G.}~\bibnamefont {Katsaros}}, \bibinfo
  {author} {\bibfnamefont {C.~M.}\ \bibnamefont {Lieber}}, \ and\ \bibinfo
  {author} {\bibfnamefont {S.}~\bibnamefont {De~Franceschi}},\ }\href {\doibase
  10.1103/PhysRevLett.109.186802} {\bibfield  {journal} {\bibinfo  {journal}
  {Phys. Rev. Lett.}\ }\textbf {\bibinfo {volume} {109}},\ \bibinfo {pages}
  {186802} (\bibinfo {year} {2012})}\BibitemShut {NoStop}%
\bibitem [{\citenamefont {Schindele}\ \emph {et~al.}(2014)\citenamefont
  {Schindele}, \citenamefont {Baumgartner}, \citenamefont {Maurand},
  \citenamefont {Weiss},\ and\ \citenamefont
  {Sch\"onenberger}}]{Schindele2014}%
  \BibitemOpen
  \bibfield  {author} {\bibinfo {author} {\bibfnamefont {J.}~\bibnamefont
  {Schindele}}, \bibinfo {author} {\bibfnamefont {A.}~\bibnamefont
  {Baumgartner}}, \bibinfo {author} {\bibfnamefont {R.}~\bibnamefont
  {Maurand}}, \bibinfo {author} {\bibfnamefont {M.}~\bibnamefont {Weiss}}, \
  and\ \bibinfo {author} {\bibfnamefont {C.}~\bibnamefont {Sch\"onenberger}},\
  }\href {\doibase 10.1103/PhysRevB.89.045422} {\bibfield  {journal} {\bibinfo
  {journal} {Phys. Rev. B}\ }\textbf {\bibinfo {volume} {89}},\ \bibinfo
  {pages} {045422} (\bibinfo {year} {2014})}\BibitemShut {NoStop}%
\bibitem [{\citenamefont {Island}\ \emph {et~al.}(2017)\citenamefont {Island},
  \citenamefont {Gaudenzi}, \citenamefont {de~Bruijckere}, \citenamefont
  {Burzur\'{\i}}, \citenamefont {Franco}, \citenamefont {Mas-Torrent},
  \citenamefont {Rovira}, \citenamefont {Veciana}, \citenamefont {Klapwijk},
  \citenamefont {Aguado},\ and\ \citenamefont {van~der Zant}}]{Island2017}%
  \BibitemOpen
  \bibfield  {author} {\bibinfo {author} {\bibfnamefont {J.~O.}\ \bibnamefont
  {Island}}, \bibinfo {author} {\bibfnamefont {R.}~\bibnamefont {Gaudenzi}},
  \bibinfo {author} {\bibfnamefont {J.}~\bibnamefont {de~Bruijckere}}, \bibinfo
  {author} {\bibfnamefont {E.}~\bibnamefont {Burzur\'{\i}}}, \bibinfo {author}
  {\bibfnamefont {C.}~\bibnamefont {Franco}}, \bibinfo {author} {\bibfnamefont
  {M.}~\bibnamefont {Mas-Torrent}}, \bibinfo {author} {\bibfnamefont
  {C.}~\bibnamefont {Rovira}}, \bibinfo {author} {\bibfnamefont
  {J.}~\bibnamefont {Veciana}}, \bibinfo {author} {\bibfnamefont {T.~M.}\
  \bibnamefont {Klapwijk}}, \bibinfo {author} {\bibfnamefont {R.}~\bibnamefont
  {Aguado}}, \ and\ \bibinfo {author} {\bibfnamefont {H.~S.~J.}\ \bibnamefont
  {van~der Zant}},\ }\href {\doibase 10.1103/PhysRevLett.118.117001} {\bibfield
   {journal} {\bibinfo  {journal} {Phys. Rev. Lett.}\ }\textbf {\bibinfo
  {volume} {118}},\ \bibinfo {pages} {117001} (\bibinfo {year}
  {2017})}\BibitemShut {NoStop}%
\bibitem [{\citenamefont {Krogstrup}\ \emph {et~al.}(2015)\citenamefont
  {Krogstrup}, \citenamefont {Ziino}, \citenamefont {Chang}, \citenamefont
  {Albrecht}, \citenamefont {Madsen}, \citenamefont {Johnson}, \citenamefont
  {Nyg{\aa}rd}, \citenamefont {Marcus},\ and\ \citenamefont
  {Jespersen}}]{Krogstrup2015}%
  \BibitemOpen
  \bibfield  {author} {\bibinfo {author} {\bibfnamefont {P.}~\bibnamefont
  {Krogstrup}}, \bibinfo {author} {\bibfnamefont {N.~L.~B.}\ \bibnamefont
  {Ziino}}, \bibinfo {author} {\bibfnamefont {W.}~\bibnamefont {Chang}},
  \bibinfo {author} {\bibfnamefont {S.~M.}\ \bibnamefont {Albrecht}}, \bibinfo
  {author} {\bibfnamefont {M.~H.}\ \bibnamefont {Madsen}}, \bibinfo {author}
  {\bibfnamefont {E.}~\bibnamefont {Johnson}}, \bibinfo {author} {\bibfnamefont
  {J.}~\bibnamefont {Nyg{\aa}rd}}, \bibinfo {author} {\bibfnamefont {C.~.~M.}\
  \bibnamefont {Marcus}}, \ and\ \bibinfo {author} {\bibfnamefont {T.~S.}\
  \bibnamefont {Jespersen}},\ }\href {\doibase 10.1038/nmat4176} {\bibfield
  {journal} {\bibinfo  {journal} {Nature Materials}\ }\textbf {\bibinfo
  {volume} {14}},\ \bibinfo {pages} {400} (\bibinfo {year} {2015})},\ \bibinfo
  {note} {article}\BibitemShut {NoStop}%
\bibitem [{\citenamefont {Taupin}\ \emph {et~al.}(2016)\citenamefont {Taupin},
  \citenamefont {Mannila}, \citenamefont {Krogstrup}, \citenamefont {Maisi},
  \citenamefont {Nguyen}, \citenamefont {Albrecht}, \citenamefont {Nyg\aa{}rd},
  \citenamefont {Marcus},\ and\ \citenamefont {Pekola}}]{Taupin2016}%
  \BibitemOpen
  \bibfield  {author} {\bibinfo {author} {\bibfnamefont {M.}~\bibnamefont
  {Taupin}}, \bibinfo {author} {\bibfnamefont {E.}~\bibnamefont {Mannila}},
  \bibinfo {author} {\bibfnamefont {P.}~\bibnamefont {Krogstrup}}, \bibinfo
  {author} {\bibfnamefont {V.~F.}\ \bibnamefont {Maisi}}, \bibinfo {author}
  {\bibfnamefont {H.}~\bibnamefont {Nguyen}}, \bibinfo {author} {\bibfnamefont
  {S.~M.}\ \bibnamefont {Albrecht}}, \bibinfo {author} {\bibfnamefont
  {J.}~\bibnamefont {Nyg\aa{}rd}}, \bibinfo {author} {\bibfnamefont {C.~M.}\
  \bibnamefont {Marcus}}, \ and\ \bibinfo {author} {\bibfnamefont {J.~P.}\
  \bibnamefont {Pekola}},\ }\href {\doibase 10.1103/PhysRevApplied.6.054017}
  {\bibfield  {journal} {\bibinfo  {journal} {Phys. Rev. Applied}\ }\textbf
  {\bibinfo {volume} {6}},\ \bibinfo {pages} {054017} (\bibinfo {year}
  {2016})}\BibitemShut {NoStop}%
\bibitem [{\citenamefont {Fisher}\ and\ \citenamefont
  {Kane}(1992)}]{Fisher1992a}%
  \BibitemOpen
  \bibfield  {author} {\bibinfo {author} {\bibfnamefont {M.~P.}\ \bibnamefont
  {Fisher}}\ and\ \bibinfo {author} {\bibfnamefont {C.}~\bibnamefont {Kane}},\
  }\href {\doibase 10.1103/PhysRevLett.68.1220} {\bibfield  {journal} {\bibinfo
   {journal} {Phys. Rev. Lett.}\ }\textbf {\bibinfo {volume} {68}},\ \bibinfo
  {pages} {1824} (\bibinfo {year} {1992})}\BibitemShut {NoStop}%
\bibitem [{\citenamefont {Kane}\ and\ \citenamefont {Fisher}(1992)}]{Kane1992}%
  \BibitemOpen
  \bibfield  {author} {\bibinfo {author} {\bibfnamefont {C.~L.}\ \bibnamefont
  {Kane}}\ and\ \bibinfo {author} {\bibfnamefont {M.~P.~A.}\ \bibnamefont
  {Fisher}},\ }\href {\doibase 10.1103/PhysRevB.46.15233} {\bibfield  {journal}
  {\bibinfo  {journal} {Phys. Rev. B}\ }\textbf {\bibinfo {volume} {46}},\
  \bibinfo {pages} {15233} (\bibinfo {year} {1992})}\BibitemShut {NoStop}%
\bibitem [{\citenamefont {Giamarchi}(2004)}]{Giamarchi2004}%
  \BibitemOpen
  \bibfield  {author} {\bibinfo {author} {\bibfnamefont {T.}~\bibnamefont
  {Giamarchi}},\ }\href@noop {} {\emph {\bibinfo {title} {Quantum Physics In
  One Dimension}}}\ (\bibinfo  {publisher} {Oxford University Press},\ \bibinfo
  {year} {2004})\BibitemShut {NoStop}%
\bibitem [{\citenamefont {Gogolin}\ \emph {et~al.}(1988)\citenamefont
  {Gogolin}, \citenamefont {Nersesyan},\ and\ \citenamefont
  {Tsvelik}}]{Gogolin1988}%
  \BibitemOpen
  \bibfield  {author} {\bibinfo {author} {\bibfnamefont {A.~O.}\ \bibnamefont
  {Gogolin}}, \bibinfo {author} {\bibfnamefont {A.~A.}\ \bibnamefont
  {Nersesyan}}, \ and\ \bibinfo {author} {\bibfnamefont {A.~M.}\ \bibnamefont
  {Tsvelik}},\ }\href@noop {} {\emph {\bibinfo {title} {Bosonization and
  strongly correlated systems}}}\ (\bibinfo  {publisher} {Cambridge},\ \bibinfo
  {year} {1988})\BibitemShut {NoStop}%
\bibitem [{\citenamefont {Takei}\ \emph {et~al.}(2013)\citenamefont {Takei},
  \citenamefont {Fregoso}, \citenamefont {Hui}, \citenamefont {Lobos},\ and\
  \citenamefont {Das~Sarma}}]{Takei13_Soft_gap}%
  \BibitemOpen
  \bibfield  {author} {\bibinfo {author} {\bibfnamefont {S.}~\bibnamefont
  {Takei}}, \bibinfo {author} {\bibfnamefont {B.~M.}\ \bibnamefont {Fregoso}},
  \bibinfo {author} {\bibfnamefont {H.-Y.}\ \bibnamefont {Hui}}, \bibinfo
  {author} {\bibfnamefont {A.~M.}\ \bibnamefont {Lobos}}, \ and\ \bibinfo
  {author} {\bibfnamefont {S.}~\bibnamefont {Das~Sarma}},\ }\href {\doibase
  10.1103/PhysRevLett.110.186803} {\bibfield  {journal} {\bibinfo  {journal}
  {Phys. Rev. Lett.}\ }\textbf {\bibinfo {volume} {110}},\ \bibinfo {pages}
  {186803} (\bibinfo {year} {2013})}\BibitemShut {NoStop}%
\bibitem [{\citenamefont {Sau}\ \emph {et~al.}(2010)\citenamefont {Sau},
  \citenamefont {Tewari}, \citenamefont {Lutchyn}, \citenamefont {Stanescu},\
  and\ \citenamefont {Sarma}}]{Sau2010}%
  \BibitemOpen
  \bibfield  {author} {\bibinfo {author} {\bibfnamefont {J.~D.}\ \bibnamefont
  {Sau}}, \bibinfo {author} {\bibfnamefont {S.}~\bibnamefont {Tewari}},
  \bibinfo {author} {\bibfnamefont {R.~M.}\ \bibnamefont {Lutchyn}}, \bibinfo
  {author} {\bibfnamefont {T.~D.}\ \bibnamefont {Stanescu}}, \ and\ \bibinfo
  {author} {\bibfnamefont {S.~D.}\ \bibnamefont {Sarma}},\ }\href {\doibase
  10.1103/physrevb.82.214509} {\bibfield  {journal} {\bibinfo  {journal} {Phys.
  Rev. B}\ }\textbf {\bibinfo {volume} {82}},\ \bibinfo {pages} {214509}
  (\bibinfo {year} {2010})}\BibitemShut {NoStop}%
\bibitem [{\citenamefont {Jespersen}\ \emph {et~al.}(2009)\citenamefont
  {Jespersen}, \citenamefont {Polianski}, \citenamefont {S{\o}rensen},
  \citenamefont {Flensberg},\ and\ \citenamefont {Nyg{\aa}rd}}]{Jespersen2009}%
  \BibitemOpen
  \bibfield  {author} {\bibinfo {author} {\bibfnamefont {T.~S.}\ \bibnamefont
  {Jespersen}}, \bibinfo {author} {\bibfnamefont {M.~L.}\ \bibnamefont
  {Polianski}}, \bibinfo {author} {\bibfnamefont {C.~B.}\ \bibnamefont
  {S{\o}rensen}}, \bibinfo {author} {\bibfnamefont {K.}~\bibnamefont
  {Flensberg}}, \ and\ \bibinfo {author} {\bibfnamefont {J.}~\bibnamefont
  {Nyg{\aa}rd}},\ }\href {\doibase 10.1088/1367-2630/11/11/113025} {\bibfield
  {journal} {\bibinfo  {journal} {New Journal of Physics}\ }\textbf {\bibinfo
  {volume} {11}},\ \bibinfo {pages} {113025} (\bibinfo {year}
  {2009})}\BibitemShut {NoStop}%
\bibitem [{\citenamefont {Albrecht}\ \emph {et~al.}(2016)\citenamefont
  {Albrecht}, \citenamefont {Higginbotham}, \citenamefont {Madsen},
  \citenamefont {Kuemmeth}, \citenamefont {Jespersen}, \citenamefont
  {Nyg{\aa}rd}, \citenamefont {Krogstrup},\ and\ \citenamefont
  {Marcus}}]{Albrecht2016}%
  \BibitemOpen
  \bibfield  {author} {\bibinfo {author} {\bibfnamefont {S.~M.}\ \bibnamefont
  {Albrecht}}, \bibinfo {author} {\bibfnamefont {A.~P.}\ \bibnamefont
  {Higginbotham}}, \bibinfo {author} {\bibfnamefont {M.}~\bibnamefont
  {Madsen}}, \bibinfo {author} {\bibfnamefont {F.}~\bibnamefont {Kuemmeth}},
  \bibinfo {author} {\bibfnamefont {T.~S.}\ \bibnamefont {Jespersen}}, \bibinfo
  {author} {\bibfnamefont {J.}~\bibnamefont {Nyg{\aa}rd}}, \bibinfo {author}
  {\bibfnamefont {P.}~\bibnamefont {Krogstrup}}, \ and\ \bibinfo {author}
  {\bibfnamefont {C.~M.}\ \bibnamefont {Marcus}},\ }\href {\doibase
  10.1038/nature17162} {\bibfield  {journal} {\bibinfo  {journal} {Nature}\
  }\textbf {\bibinfo {volume} {531}},\ \bibinfo {pages} {206 EP} (\bibinfo
  {year} {2016})}\BibitemShut {NoStop}%
\bibitem [{\citenamefont {Haldane}(1981)}]{Haldane1981}%
  \BibitemOpen
  \bibfield  {author} {\bibinfo {author} {\bibfnamefont {F.~D.~M.}\
  \bibnamefont {Haldane}},\ }\href {\doibase 10.1103/PhysRevLett.47.1840}
  {\bibfield  {journal} {\bibinfo  {journal} {Phys. Rev. Lett.}\ }\textbf
  {\bibinfo {volume} {47}},\ \bibinfo {pages} {1840} (\bibinfo {year}
  {1981})}\BibitemShut {NoStop}%
\bibitem [{\citenamefont {Bortolin}\ \emph {et~al.}()\citenamefont {Bortolin},
  \citenamefont {Lobos},\ and\ \citenamefont {Iucci}}]{Bortolin2019}%
  \BibitemOpen
  \bibfield  {author} {\bibinfo {author} {\bibfnamefont {T.}~\bibnamefont
  {Bortolin}}, \bibinfo {author} {\bibfnamefont {A.}~\bibnamefont {Lobos}}, \
  and\ \bibinfo {author} {\bibfnamefont {A.}~\bibnamefont {Iucci}},\
  }\href@noop {} {\ }\bibinfo {note} {In preparation}\BibitemShut {NoStop}%
\bibitem [{\citenamefont {Wu}\ \emph {et~al.}(2006)\citenamefont {Wu},
  \citenamefont {Bernevig},\ and\ \citenamefont
  {Zhang}}]{Wu06_Helical_liquid_and_the_QSHE}%
  \BibitemOpen
  \bibfield  {author} {\bibinfo {author} {\bibfnamefont {C.}~\bibnamefont
  {Wu}}, \bibinfo {author} {\bibfnamefont {B.~A.}\ \bibnamefont {Bernevig}}, \
  and\ \bibinfo {author} {\bibfnamefont {S.-C.}\ \bibnamefont {Zhang}},\ }\href
  {\doibase 10.1103/PhysRevLett.96.106401} {\bibfield  {journal} {\bibinfo
  {journal} {Phys. Rev. Lett.}\ }\textbf {\bibinfo {volume} {96}},\ \bibinfo
  {pages} {106401} (\bibinfo {year} {2006})}\BibitemShut {NoStop}%
\end{thebibliography}
%

\end{document}